\documentclass[aps,prb,showpacs]{revtex4}
\usepackage{graphicx,amsmath,amssymb,color}
\usepackage{float}
\begin{document}
\title{ Effect of Temperature and Doping on Plasmon Excitations for an Encapsulated
Double-Layer Graphene Heterostructure}
\author{  Godfrey Gumbs$^{1,2}$, Dipendra Dahal$^1$, and Antonios Balassis$^3$  }
\affiliation{$^{1}$Department of Physics and Astronomy, Hunter College of the
City University of New York, 695 Park Avenue, New York, NY 10065, USA\\
$^{2}$ Donostia International Physics Center (DIPC),
Paseo de Manuel Lardizabal 4, 20018 San Sebasti\'an/Donostia, Spain\\
$^{3}$Department of Physics \& Engineering Physics, Fordham  University, 441 East Fordham Road, Bronx, NY 10458,  USA }

\date{\today}

\begin{abstract}
We perform a comprehensive analysis of the spectrum of graphene plasmons which arise when a pair of
sheets are confined between conducting materials.  The associated enhanced  local fields
may be employed in the manipulation of light on the nanoscale by adjusting
the separation between the graphene layers, the energy band gap as well as the
concentration of charge carriers in the conducting media surrounding the
two-dimensional (2D)  layers.   We present a theoretical formalism, based on 
the calculation of the surface response function,  for  determining
 the plasmon spectrum of an encapsulated pair of 2D  layers  and apply it to graphene.
We solve the coupled equations involving the  continuity of the electric potential and 
discontinuity of the electric field at the interfaces separating the constituents of the hybrid 
structure. We have compared the plasmon modes for encapsulated gapped and gapless graphene.
The associated nonlocal graphene plasmon spectrum coupled to the ``sandwich"
system   show a linear acoustic plasmon mode as well as a low-frequency mode corresponding to
in-phase oscillations of the adjacent  2D charge densities. 
These calculations are  relevant to the study of energy 
transfer via plasmon excitations when graphene is confined by a pair of thick conducting materials.

\end{abstract}

\vskip 0.2in

\pacs{73.21.-b, 71.70.Ej, 73.20.Mf, 71.45.Gm, 71.10.Ca, 81.05.ue}

\medskip
\par
\maketitle

\section{Introduction}
\label{sec1}

Many researchers have been devoting a great deal of effort to exploit the unique transport and optical
properties of graphene. In particular, an area of much interest to both experimentalists and theoreticians 
has been the study of plasmon excitations under various conditions 
of temperature and doping concentrations.   There have been many recent works focused on the 
study of these plasmon modes in graphene when it is free standing\cite{wunsch}, lying on
a substrate\cite{GG,revised1} or encapsulated by two conducting materials.\cite{GG2}  In this paper, we  investigate   
 the way in which the plasmon mode excitations\cite {revised2} for a pair of graphene layers are affected by encapsulating conductors
which are coupled non locally to the two-dimensional (2D) layers.      
 
 \medskip
 \par

The unusual properties of free-standing graphene may be attributed to Bloch states
in the corners of the hexagonal Brillouin zone of this 2D honeycomb crystal lattice.
For example, the Dirac fermions arising from this energy band structure  lead to
strongly enhanced and confined local fields through dipole-dipole coupling \cite{Cox}.
But, recently, novel properties have been predicted when graphene electrically interacts with
a nearby metallic substrate separated by a thin insulator
\cite{SS13,SS14,SS15,encaps1,encaps2,encaps3,encaps4,encaps5,encaps6,encaps7,encaps8,encaps9,encaps10}.
The graphene-insulator-metal plasmons have exhibited both a linear dispersion
mode \cite{SS15,SS18} (a so-called acoustic plasmon) in the terahertz (THz) spectral regime
and a plasmon mode which is 2D in nature (square root of the wave vector).
Such plasmon excitations are important for improving THz sensing, signal processing on electro-optic 
modulation\cite{revised3} and communication
technologies \cite{SS2,SS3,SS21}.

\medskip
\par

 Reportedly,  graphene encapsulated in hexagonal boron-nitride (hBN) displays
an anomalous Hall effect at room temperature which may be interpreted as ballistic transport
on a micrometer scale over a wide range of carrier concentration. The encapsulation makes
graphene virtually protected from  its surroundings but at the same time allows the use
of hBN as a top gate dielectric.  More generally, the properties of encapsulated graphene are 
currently being actively investigated due to recent advances in device fabrication 
techniques \cite{encaps1,encaps2,encaps3,encaps4,encaps5,encaps6,encaps7,encaps8,encaps9,encaps10,Gong,Kamat}.
One reason for the attention being paid to these heterostructures is the observed
improvements in the electrical conductance of graphene interconnection when there is complete encapsulation
by boron-nitride \cite{encaps10,XXX4,YYY1,YYY2,YYY5,YYY6}. Another is that when graphene interacts with a 
substrate such as SiC, hBN, graphite\cite{substrate1,substrate2,substrate3}, an energy band gap opens up and 
the plasmon dispersion relation is modified.

  \medskip
  \par

 The combination of a 2D layer interacting with a substrate presents
theoreticians with a challenge to model the structure as well as to formulate the
problem and eventually obtain a dispersion equation for the plasmon excitations.
There already is a formulation for graphene plasmons on a substrate  \cite{GG,ONB,NJMH}
 as well as when encapsulated by a pair of thick conductors \cite{GG2}. These works were geared
to help explain the reported relevant experimental data \cite{Pol1,Pol2,Pol3,Pol4,Pol5}.  
Our present investigation showed that the plasmons of an encapsulated  double-layer graphene 
heterostructure  are  different in  dispersive nature  from the case when there are two graphene 
layers interacting with a single substrate.

\medskip
\par
 In this paper, we present a formalism which yields the surface response function for an encapsulated
pair of 2D layers.  We take into account the presence of a dielectric medium separating
the 2D layer from the substrate. Furthermore, the top encapsulating material could have finite thickness
as shown schematically in Fig.~\ref{FIG:1}.  In general, this method is capable of obtaining the 
plasmon dispersion relation for an arbitrary number of  encapsulated layers embedded within a background 
dielectric material which ensures that there is no electron tunneling in between the layers. In addition, 
our method can be  extended to the case when a protective  coating is placed over the top substrate. 
The surface response function is calculated in the vicinity of the surface and our central formula
yields various already published results.\cite{GG,Bopersson,Dassarma}   
This response function, thereby obtained is used to determine the plasmon modes which are exhibited as 
density plots. Our method includes the effect of dielectric media non locally coupled which was not investigated
by Badalyan and Peeters \cite{Badalyan} who treated the effect of the surrounding dielectric media locally.
For further clarification, we compare the resulting calculated  plasmon spectra for encapsulated gapless and gapped graphene 
in the presence and absence of the background dielectric. Besides this, we  investigated the effect of temperature on
the plasmon dispersion.  A similar effect can be calculated for transition metal dichalcogenide monolayers, 
modeling them as we do for gapped graphene\cite{tmd1,tmd2,tmd3,tmd4}.  On the other hand, the structure  
which we have examined here may play an important role in increasing the mobility of electrons in graphene 
and other 2D structure to improve the performance of nanoelectronic and spintronic devices.  The plasmon mode 
excitation of the structure could also be used as an illumination source for imaging a sample\cite{revised4} or 
in optoelectronic devices as a sensor.

\medskip
\par

We now outline the rest of this paper as follows.  In Sec.~\ref{sec2},
we present a detailed description  of our method for calculating  the surface response function for
a pair of 2D layers sandwiched between two conducting substrates with arbitrary separation. The calculated 
plasmon spectra for chosen energy band gap, carrier doping values and temperature effect are presented 
in Sec.~\ref{sec3}. We conclude our paper with a discussion of the highlights of our calculations
in Sec.~\ref{sec4}.

\section{  Theoretical   Formalism}
\label{sec2}

Let us consider a structure as shown in Fig.~\ref{FIG:1} where a 2D layer  is placed on top of a 
substrate of thickness $\ell_1$ with frequency-dependent dielectric function $\epsilon_1(\omega)$.
Below this, there is a buffer layer between $z=\ell_1$ and $z=\ell_4$ with  background dielectric 
constant $\epsilon_b$  and a pair of 2D layers are embedded at $z=\ell_2$ and at $z=\ell_3$.    
This entire structure is now placed  over a thick conducting substrate of dielectric 
function $\epsilon_2(\omega)$  which extends from $z=\ell_4$ to infinity.  
For this structure we determine the surface response function which is used to calculate 
the plasmon frequencies. 

\medskip
\par
The electrostatic potential in the vicinity of the surface for the heterostructure is written as \cite{GGC}    

\begin{equation}
\phi_<(z)=e^{-q_{\parallel}z}-g(q_{\parallel},\omega)e^{q_{\parallel}z}\ ,  z \lessapprox 0 
\end{equation}
which introduces the surface response function $g(q_{\parallel},\omega)$ which is the
object  of our calculation.  To proceed further, we express the potential 
solutions of Poisson's equation in the various regions depicted in Fig.~\ref{FIG:1} as linear combinations
of $e^{\pm q_{\parallel}z}$ by making use of the translational invariance parallel to the $xy$-planar interfaces.

\medskip
\par

Referring to Fig.\ \ref{FIG:1},  we write
the potential functions in the different regions as follows:

\begin{eqnarray}
\phi_>(z) &=& a_1e^{q_{\parallel}z}+b_1e^{-q_{\parallel}z}\ , \ \ 0\leq z\leq \ell_1
\nonumber\\
\phi_{1<}(z)&=& t_1e^{q_{\parallel}z}+r_1e^{-q_{\parallel}z}\ , \ \ \ell_1\leq z\leq \ell_2
\nonumber\\
\phi_{1>}(z) &=& t_2e^{q_{\parallel}z}+r_2e^{-q_{\parallel}z} \ , \ \ \ell_2\leq z\leq \ell_3
\nonumber\\
\phi_{2<}(z) &=& f_1e^{q_{\parallel}z}+k_1e^{-q_{\parallel}z} \ , \ \ \ell_3\leq z\leq \ell_4
\nonumber\\
\phi_{2>}(z) &=& k_2e^{-q_{\parallel}z} \ , \ \   z\geq \ell_4
\end{eqnarray}
where $a_1$, $b_1$, $t_1$, $r_1$, $t_2$, $r_2$, $f_1$, $k_1$ and $k_2$ are  independent of $z$.
These are readily determined by using the continuity of the potential and the discontinuity 
of the electric field across the interfaces.  We assume that the 2D layer at $z=0$ 
has induced surface charge density  $\sigma_1$ and that at $z=\ell_2$ and $z=\ell_3$ has  
charge density $\sigma_2$ and $\sigma_3$, respectively.  Also, using linear response theory, we have
$\sigma_1=\chi_1 \phi_<(0)$,  $\sigma_2=\chi_2\phi_{1<}(\ell_2)$  and $\sigma_3=\chi_3\phi_{2<}(\ell_3)$, 
where $\chi_1$, $\chi_2$  and $\chi_3$ are 2D susceptibilities.  After some algebra, we can obtain the 
surface response function for the structure shown in Fig.\ \ref{FIG:1} with the cap 2D layer.
However, here we are concerned with a heterostructure without the protective layer at $z=0$ since this 
simplifies our calculation and which is reasonable to neglect if $\ell_1$ is large.
For this,  we set $\sigma_1=0$ ,  $\ell_1=d_1$, $\ell_2=d_1+d_2$, $\ell_3=d_1+d_2+d_3$ and $ \ell_4=d_1+d_2+d_3+d_4$. 
In this notation, we obtain $g(q_\parallel,\omega)$ for {\em two\/} encapsulated sheets.  
Setting $d_1$, $d_2=d/2$, $d_3=d$, $d_4=d/2$, we have
\begin{equation}
g(q_{\parallel},\omega)=\frac{{\cal N}(q_{\parallel},\omega)}{{\cal  D}(q_{\parallel},\omega)} \ . 
\label{general}
\end{equation}
Here,
\begin{eqnarray}
{\cal N}(q_{\parallel},\omega)&&=2q_{\parallel}^2\epsilon_0^2\epsilon_b^2\left[\sinh(q_{\parallel}d_1)N_1+\cosh(q_{\parallel}d_1)\epsilon_1N_2\right]\nonumber\\&&-2\sinh(q_{\parallel}d)N_3N_4\chi_2\chi_3+q_{\parallel}\epsilon_0\epsilon_b\left[\sinh(q_{\parallel}d)N_5\epsilon_b(\chi_2-\chi_3)-\sinh(2q_{\parallel}d)N_6\epsilon_b(\chi_2+\chi_3)\right.\nonumber\\&&+\left.\cosh(2q_{\parallel}d)N_7(\chi_2+\chi_3)+\cosh(q_{\parallel}d)N_8(\chi_2+\chi_3)\right]
\label{HN1}
\end{eqnarray}
with $\epsilon_0$ denoting the permittivity  of free space, for convenience we suppress the 
frequency-dependence of some quantities, 

\begin{equation}
N_1=\sinh (2 q_{\parallel}d) \left(\text{$\epsilon $}_1^2 \text{$\epsilon $}_2-\text{$\epsilon $}_b^2\right)+\text{$\epsilon $}_b \left(\text{$\epsilon $}_1^2-\text{$\epsilon $}_2\right) \cosh (2 q_{\parallel}d),
\end{equation}

\begin{equation}
N_2=\left(\text{$\epsilon $}_b^2-\text{$\epsilon $}_2\right) \sinh (2 q_{\parallel}d)+(\text{$\epsilon $}_2-1) \text{$\epsilon $}_b \cosh (2 q_{\parallel}d),
\end{equation}

\begin{equation}
N_3=\text{$\epsilon $}_2 \sinh \left(\frac{q_{\parallel}d}{2}\right)+\text{$\epsilon $}_b \cosh \left(\frac{q_{\parallel}d}{2}\right),
\end{equation}

\begin{equation}
N_4=\text{$\epsilon $}_b \cosh \left(\frac{q_{\parallel}d}{2}\right) \left[\sinh (q_{\parallel}d_1)-\text{$\epsilon $}_1 \cosh (q_{\parallel}d_1)\right]+\text{$\epsilon $}_1 \sinh \left(\frac{q_{\parallel}d}{2}\right) \left[\cosh (q_{\parallel}d_1)-\text{$\epsilon $}_1 \sinh (q_{\parallel}d_1)\right],
\end{equation}

\begin{equation}
N_5=\left(\text{$\epsilon $}_1^2+\text{$\epsilon $}_2\right) \sinh (q_{\parallel}d_1)-\text{$\epsilon $}_1 (\text{$\epsilon $}_2+1) \cosh (q_{\parallel}d_1),
\end{equation}

\begin{equation}
N_6=\left.\left(\text{$\epsilon $}_1^2-\text{$\epsilon $}_2\right) \sinh (q_{\parallel}d_1)+\text{$\epsilon $}_1 (\text{$\epsilon $}_2-1) \cosh (q_{\parallel}d_1)\right),
\end{equation}

\begin{equation}
N_7=\sinh (q_{\parallel}d_1) \left(\text{$\epsilon $}_b^2-\text{$\epsilon $}_1^2 \text{$\epsilon $}_2\right)+\text{$\epsilon $}_1 \left(\text{$\epsilon $}_2-\text{$\epsilon $}_b^2\right) \cosh (q_{\parallel}d_1),
\end{equation}

\begin{equation}
N_8=\sinh (q_{\parallel}d_1) \left(\text{$\epsilon $}_1^2 \text{$\epsilon $}_2+\text{$\epsilon $}_b^2\right)-\text{$\epsilon $}_1 \left(\text{$\epsilon $}_2+\text{$\epsilon $}_b^2\right) \cosh (q_{\parallel}d_1),
\end{equation}
and
\begin{eqnarray}
{\cal  D}(q_{\parallel},\omega)&&=2q_\parallel^2\epsilon_0^2\epsilon_b^2\left[\cosh(q_{\parallel}d_1)\epsilon_1D_1+\sinh(q_{\parallel}d_1)D_2\right]\nonumber\\&&+2\sinh(q_{\parallel}d)D_3D_4\chi_2\chi_3-q_{\parallel}\epsilon_0\epsilon_b\left[\sinh(q_{\parallel}d)D_5\epsilon_b(\chi_2-\chi_3)+\sinh(2q_{\parallel}d)D_6\epsilon_b(\chi_2+\chi_3)\right.\nonumber\\&&-\left.\cosh(q_{\parallel}d)D_7(\chi_2+\chi_3)+\cosh(2q_{\parallel}d)D_8(\chi_2+\chi_3)\right]
\label{HN2}
\end{eqnarray}
with

\begin{equation}
D_1=\left(\text{$\epsilon $}_2+\text{$\epsilon $}_b^2\right) \sinh (2q_{\parallel}d)+(\text{$\epsilon $}_2+1) \text{$\epsilon $}_b \cosh (2 q_{\parallel}d),
\end{equation}

\begin{equation}
D_2=\sinh (2 q_{\parallel}d) \left(\text{$\epsilon $}_1^2 \text{$\epsilon $}_2+\text{$\epsilon $}_b^2\right)+\text{$\epsilon $}_b \left(\text{$\epsilon $}_1^2+\text{$\epsilon $}_2\right) \cosh (2 q_{\parallel}d),
\end{equation}
\begin{equation}
D_3=\text{$\epsilon $}_2 \sinh \left(\frac{q_{\parallel}d}{2}\right)+\text{$\epsilon $}_b \cosh \left(\frac{q_{\parallel}d}{2}\right),
\end{equation}

\begin{equation}
D_4=\text{$\epsilon $}_b \cosh \left(\frac{q_{\parallel}d}{2}\right) \left[\text{$\epsilon $}_1 \cosh (q_{\parallel}d_1)+\sinh (q_{\parallel}d_1)\right]+\text{$\epsilon $}_1 \sinh \left(\frac{q_{\parallel}d}{2}\right) \left[\text{$\epsilon $}_1 \sinh (q_{\parallel}d_1)+\cosh (q_{\parallel}d_1)\right],
\end{equation}

\begin{equation}
D_5=\left(\text{$\epsilon $}_2-\text{$\epsilon $}_1^2\right) \sinh (q_{\parallel}d_1)+\text{$\epsilon $}_1 (\text{$\epsilon $}_2-1) \cosh (q_{\parallel}d_1),
\end{equation}

\begin{equation}
D_6=\left(\text{$\epsilon $}_1^2+\text{$\epsilon $}_2\right) \sinh (q_{\parallel}d_1)+\text{$\epsilon $}_1 (\text{$\epsilon $}_2+1) \cosh (q_{\parallel}d_1),
\end{equation}

\begin{equation}
D_7=\sinh (q_{\parallel}d_1) \left(\text{$\epsilon $}_1^2 \text{$\epsilon $}_2-\text{$\epsilon $}_b^2\right)+\text{$\epsilon $}_1 \left(\text{$\epsilon $}_2-\text{$\epsilon $}_b^2\right) \cosh (q_{\parallel}d_1),
\end{equation}

\begin{equation}
D_8=\sinh (q_{\parallel}d_1) \left(\text{$\epsilon $}_1^2 \text{$\epsilon $}_2+\text{$\epsilon $}_b^2\right)+\text{$\epsilon $}_1 \left(\text{$\epsilon $}_2+\text{$\epsilon $}_b^2\right) \cosh (q_{\parallel}d_1),
\end{equation}

where, $\epsilon_1(\omega)=\epsilon_2(\omega)=1- \omega_p^2/\omega^2$ and $\epsilon_b$ is constant.

\medskip
\par

It is of interest to note that Persson\cite{Bopersson} calculated the surface response function for 
a 2D sheet lying on top of a substrate with dielectric function $\epsilon_2(\omega)$ with vacuum on the other side 
as  

\begin{equation}
g_{\text{P}}(q_{\parallel},\omega)=1-\frac{2}{ 1+\epsilon_2-\frac{\chi_3}{q_{\parallel}\epsilon_0}}
\end{equation}
which is readily recovered from our general result in Eq. (~\ref{general})  
by replacing $d_1=0$, $d=0$ with $\chi_2=0$.
Similarly, Hwang and Das Sarma \cite{Dassarma} obtained the plasmon dispersion equation for a pair of
free standing 2D layers with a chosen separation between them.  When we make the substitution  
$\epsilon_1(\omega)=1$, $\epsilon_2(\omega)=1$ and $\epsilon_b=1$ in Eq.\ (\ref{general}), we obtain

\begin{equation}
g_{\text{free-standing}}(q_{\parallel},\omega)=\frac{e^{ -q_{\parallel}\left(3d+2 d_1\right)} 
\left[\chi _2 \chi _3 \left(e^{2 q_{\parallel}d }-1\right)-2 q_{\parallel} \epsilon _0 
\left(\chi _2 e^{2  q_{\parallel}d}+\chi _3\right)\right]
}{4 q_{\parallel}^2 \epsilon_0^2 \left[\big(1-\frac{\chi_2}{2 q_{\parallel}\epsilon_0}
\big)\big(1-\frac{\chi_3}{2 q_{\parallel} \epsilon_0}\big)-
\frac{\chi_2}{2 q_{\parallel} \epsilon_0}\frac{\chi_3}{2 q_{\parallel} \epsilon_0}e^{-2 q_{\parallel} d}\right]}\ .
\label{HDS}
\end{equation}
The zeros of the denominator in Eq. (~\ref{HDS})   correspond to the plasmon poles and the resulting 
dispersion equation agrees with that in Ref.\ [\onlinecite{Dassarma}].

\medskip
\par

The dispersion equation for plasmon excitations when a 2D  layer is located at some distance from the 
surface of a thick conducting substrate, as considered by Gumbs, et al.\cite{GG} can be successfully deduced 
from Eq.\ (\ref{general}) by substituting  $d_1=0$, $\chi_3=0$,  $\epsilon_b=1$ and $\epsilon_2=\epsilon$. 
The surface response function for this case is 
\begin{equation}
g_{\text{2D-substrate}}(q_{\parallel},\omega)=\frac{\left(\epsilon-1\right) 
\left(2 q_{\parallel} \epsilon _0+\chi _2\right)-\chi _2 \left(\epsilon+1\right) 
e^{3  q_{\parallel}d}}{2q_{\parallel}\epsilon_0(\epsilon+1)e^{4 q_{\parallel}d}
\left[1-\frac{\chi_2}{2 q_{\parallel} \epsilon_0}
\big(1+\frac{1-\epsilon}{1+\epsilon}e^{-2 q_{\parallel}\frac{3d}{2}}\big)\right]}
\label{PRB1}
\end{equation}
from which the dispersion equation is obtained by setting the  factor in parenthesis in the denominator 
of this equation equal to zero and this agrees with the result in Ref.~[\onlinecite{GG}].  We emphasize that
in this special case, the distance from the conducting substrate to the 2D layer is $\ 3d/2$.

\medskip
\par

Next, we turn to the situation when there is a single 2D layer which is sandwiched between 
a conducting material of finite thickness on one side and by a semi-infinite conductor on
the other side.
The surface response function of this structure can be deduced from Eq.~(\ref{general})     
by setting $\chi_2=0$ and $d=0$. Then the surface response function obtained is as follows:

\begin{equation}
g_{\text{single-layer}}(q_{\parallel},\omega)=1-\frac{2\left[\big(\epsilon_1+\epsilon_2-\frac{\chi_3}{q_{\parallel}\epsilon_0}
\big)+\big(\epsilon_1-\epsilon_2+\frac{\chi_3}{q_{\parallel}\epsilon_0}\big)e^{-2q_{\parallel}d_1}
\right]}{\left[(1+\epsilon_1)(\epsilon_1+\epsilon_2-
\frac{\chi_3}{q_{\parallel}\epsilon_0})-
(\epsilon_1-1)(\epsilon_1-\epsilon_2+\frac{\chi_3}{q_{\parallel}\epsilon_0})e^{-2q_{\parallel}d_1}\right]}\ .
\end{equation}
Equating the denominator of the second term to zero, we obtain the plasmon dispersion equation for one encapsulated layer
with susceptibility $\chi_3$. 
Furthermore, if one sets $\chi_2$ and $\chi_3$ equal to zero in Eqs.\ (\ref{HN1}) and (\ref{HN2}), i.e., all the 2D
layers in Fig.\ \ref{FIG:1}  are removed, we obtain an expression for the surface response function of a thick slab 
consisting of regions with different dielectric materials.      In particular, one may then also set $\epsilon_2=1$,
$d=0$, $d_1=L$  and $\epsilon_1=\epsilon$ to obtain 
 the surface response function  for a film of thickness $L$ given by 

\begin{equation}
g_{{\mbox film}}(q_{\parallel},\omega)=2\frac{g_\infty(q_{\parallel},\omega)}{e^{q_\parallel L}-g_\infty^2(q_{\parallel},\omega)e^{-q_\parallel L}}
\sinh (q_\parallel L)
\label{ADD3}
\end{equation}
where $g_\infty(q_{\parallel},\omega)=(\epsilon-1)/(\epsilon+1)$

\medskip
\par

\medskip
\par 

The preceding calculations clearly confirm that our result in Eq.\ (\ref{general}) is very useful and can be 
employed for a variety of structures. As we pointed out above, we may extend our calculations to include 
the influence of a protective top layer, as schematically demonstrated in Fig.~\ref{FIG:1}.
The plasmon frequency of an encapsulating substrate such as $\text{Bi}_2\text{Se}_3$\cite{Apolitano} is calculated  using
$\omega_p=\sqrt{ n_{3D}e^2/(\epsilon_b\epsilon_0m^\ast)}$=270 meV for $n_{3D} \approx 8.2 \times 10^{18}\text{cm}^{-3}$ 
and $m^\ast=0.15 m_e$ representing  the charge density of the conducting substrate and the effective mass of an electron 
given in terms of $m_e$, the free electron rest mass, respectively. Consequently,  the background dielectric constant of  a conductor can vary significantly from $\epsilon_b=200-500$ for barium or strontium titanate, to many orders of magnitude greater for clean, copper-based metals and the plasmon frequency may vary over a wide range up to $10^{12}$ Hz, which corresponds to a few meV. On the other hand, the zero temperature doping(Fermi energy) in graphene could be estimated as $E_F=\hbar v_fk_F=\hbar v_f \sqrt{\pi n_{2D}}=0.04$eV \cite{DasLi} for two dimensional charge density, $n_{2D}=10^{15}\text{m}^{-2}$ and Fermi wave vector, $k_F^{\Delta}=\sqrt{\mu^2-\Delta^2}/\hbar v_F$ with $v_F$, $\mu$ and $\Delta$ as Fermi velocity, chemical potential and half band gap respectively.
Thus, both quantities are of same order of magnitude, and using $\mu$ as the unit for frequency is reasonable.
Now, in order to determine the plasmon modes for a pair of encapusulated 2D graphene layers numerically,
the denominator, ${\cal D}(q_{\parallel},\omega,T)$ in Eq.\ (\ref{general}) is equated to zero with 
$\chi_2=\chi_3=-e^2\Pi^{(0)}_{2D}(q_{\parallel},\omega,T)$, where $\Pi^{(0)}_{2D}(q_{\parallel},\omega,T)$ 
is the polarization function of graphene with

\begin{equation}
\Pi^{(0)}_{2D}(q_\parallel,\omega,T) = \frac{1}{2 \pi^2}
\int d^2 {\bf k} \sum\limits_{s, s' = \pm 1}
\left\{1+ s s'\frac{ \hbar^2v_F^2({ \bf k}+ { \bf q})\cdot { \bf k}+\Delta^2} { E_k\,E_{|{\bf k+q}|}} \right\}
\frac{f_0(s E_{k}-\mu,T)-f_0(s' E_{|{\bf k+q}|}-\mu,T)}{s E_k - s' E_{|{\bf k+q}|}
-\hbar (\omega + i 0^+)} \ ,
\label{eqn33}
\end{equation}
where,  $f_0(s E_{k}-\mu,T)=(1+e^{(sE_k-\mu)/k_BT})^{-1}$ is the Fermi-Dirac distribution function for subband energy, 
$s E_{k}=s\sqrt{(\hbar v_F k)^2+\Delta^2}$  with $s=\pm 1$ and $T$ is  temperature of the system.

To investigate the effect of encapsulation  on gapless and gapped graphene, we employ  the polarization function of
Wunsch\cite{wunsch} and Pyatkovskiy\cite{Pyatkovskiy} respectively whereas to see the effect of temperature on the 
plasmon modes, we make use of the results given in Ref.~[\onlinecite{Ramezanali}].

\section{Numerical Results and Discussion} 
\label{sec3}
In Figs.~\ref{FIG:2} and \ref{FIG:3}, we present our numerical results for nonlocal plasmon excitations of a pair of gapless and 
gapped graphene layers encapsulated by dielectric materials above and below them. On one side, the  dielectric has finite thickness 
and on the other side, there is a semi-infinite material. Figure \ref{FIG:2} portrays the plasmon modes  for encapsulated double 
layer gapless graphene when the double layer system is in vacuum(left panel) and in background dielectric with $\epsilon_b=2.4$(right panel). In total, four plasmon excitation modes are obtained, two of which originating 
from the origin are referred as the acoustic plasmon (AP), which is linear in the wave vector at long wavelengths 
and has the lowest frequency, and a mode lying above it which is 2D-like that we refer to as the optical plasmon (OP).   
The other two plasmon modes originating from the bulk plasma frequency $\omega_p$ and the surface plasmon frequency 
$\omega_p/\sqrt{2}$ are labeled as  the upper hybrid plasmon (UHP) and the lower hybrid plasmon (LHP), respectively.
All plasmon modes survive in the longer wavelength region and get Landau damped as the mode enters the particle-hole 
mode region at shorter wavelengths.  Comparison of the results on the left-hand side of Fig.\ \ref{FIG:2} with 
those on the right shows that  due to embedding of the 2D layers in the dielcectric background the UHP mode 
flattens out and becomes less dispersive. In fact, the AP mode  is noticeably closer to the particle hole mode 
region and the  range of wave vector for which this  {\em not\/} Landau damped is reduced. Besides this, the 
embedding of graphene layers in a dielectric background leads to the separation of the AP and OP mode in the 
shorter wavelength region  where they were merged when there was no background dielectric screening.

\medskip
\par

Figure \ref{FIG:3} presents numerical results for encapsulated double-layer gapped graphene.  The results
in the left panel of these plots are for a double-layer gapped graphene system lying in vacuum and the 
right panel figures are for cases when the  layers are embedded in a background dielectric with constant 
$\epsilon_b$.  As in the case of gapless graphene, when gapped graphene layers are embedded in a dielectric,
we observe four plasmon modes for which an AP and OP mode originate 
from the origin and the other two hybrid plasmon modes, the LHP and UHP stem from the surface and bulk 
plasma frequencies of the substrate. A corresponding panel comparison of Fig. \ref{FIG:2} with Fig. \ref{FIG:3} 
clearly shows that due to the presence of the energy band gap, the particle-hole mode excitation region splits 
into two parts.  The reason is due to an increase in the energy required for an 
electron to transfer form the valence to the conduction band.  
When the band gap is small of range $\Delta=0.2\hbar\omega_p$, the splitting is not large enough 
for the plasmon modes to enter.   However, as the band gap is increased, the splitting widens and  can
allow the plasmon modes originating from the bulk and surface plasma frequencies of the substrate not to undergo
any Landau damping over a wide range of wave numbers. As the band gap continues to be increased, the two hybrid 
modes merge into one for shorter wavelengths. Beside this, we observe that when the 2D layers are embedded in a 
background dielectric material (right-hand side panels in the  figures)  the oscillator strengths of all modes 
become smaller and the AP mode lies closer to the particle-hole excitation region where it is Landau damped. 
The AP and OP modes which are degenerate in the shorter wave length region now become non-coincident when the 
pair of graphene layers is embedded in the dielectric background.

\medskip
\par

Our numerical calculations for the effect of temperature on the heterostructure, displayed in Fig. \ref{FIG:4},
encapsulating gapless graphene reveals that the behavior of the plasmon spectra may be affected 
non uniformly with respect to wave vector and frequency  even at a chosen temperature. At zero temperature,  
the Fermi energy is equal to $0.04 eV$, the corresponding Fermi temperature is $T_F \approx 450$ K
and  the Fermi wave vector  is $k_F \approx 10^7m^{-1}$.  We may choose  the plasma frequency 
$\omega_p \approx 10^{13} Hz$ so that we could take  $T=0.5 T_F$ in our calculations. At this temperature,
the electron is thermally excited and the transition between the valence band and conduction band occurs
continuously causing the boundary of the single-particle excitations to smear out which  is a distinct 
difference from that seen in Figs. \ref{FIG:2} and  \ref{FIG:4} as an effect due to temperature. This 
effect due to smearing resulted in the removal of sharp boundary of the particle hole excitation region in wave 
vector-frequency space  for which there is none at absolute zero. Due to this, the 
two hybrid modes originating from the vicinity of $\omega_p$ seem to be affected the most. 
The UHP mode dies off soon  after it emerges with the bulk frequency in the long wavelength regime
whereas the LHP mode survives over a wider range of wave vector and  decays as $q_\parallel$ is increased.
The AP and OP modes  survive for longer wavelength and decay in the shorter wavelength region.  
Also, another distinct behavior observed, when the double layer of graphene   
is embedded in a dielectric material is that the AP mode moves closer to the particle-hole  
region and becomes decayed at longer wavelengths. In addition, the AP mode and OP mode become non-degenerate 
in the shorter wavelength regime.

\section{  Concluding Remarks}
\label{sec4}

\begin{figure}[t]
\centering
\includegraphics[width=8.4in, height=7.1in,keepaspectratio]{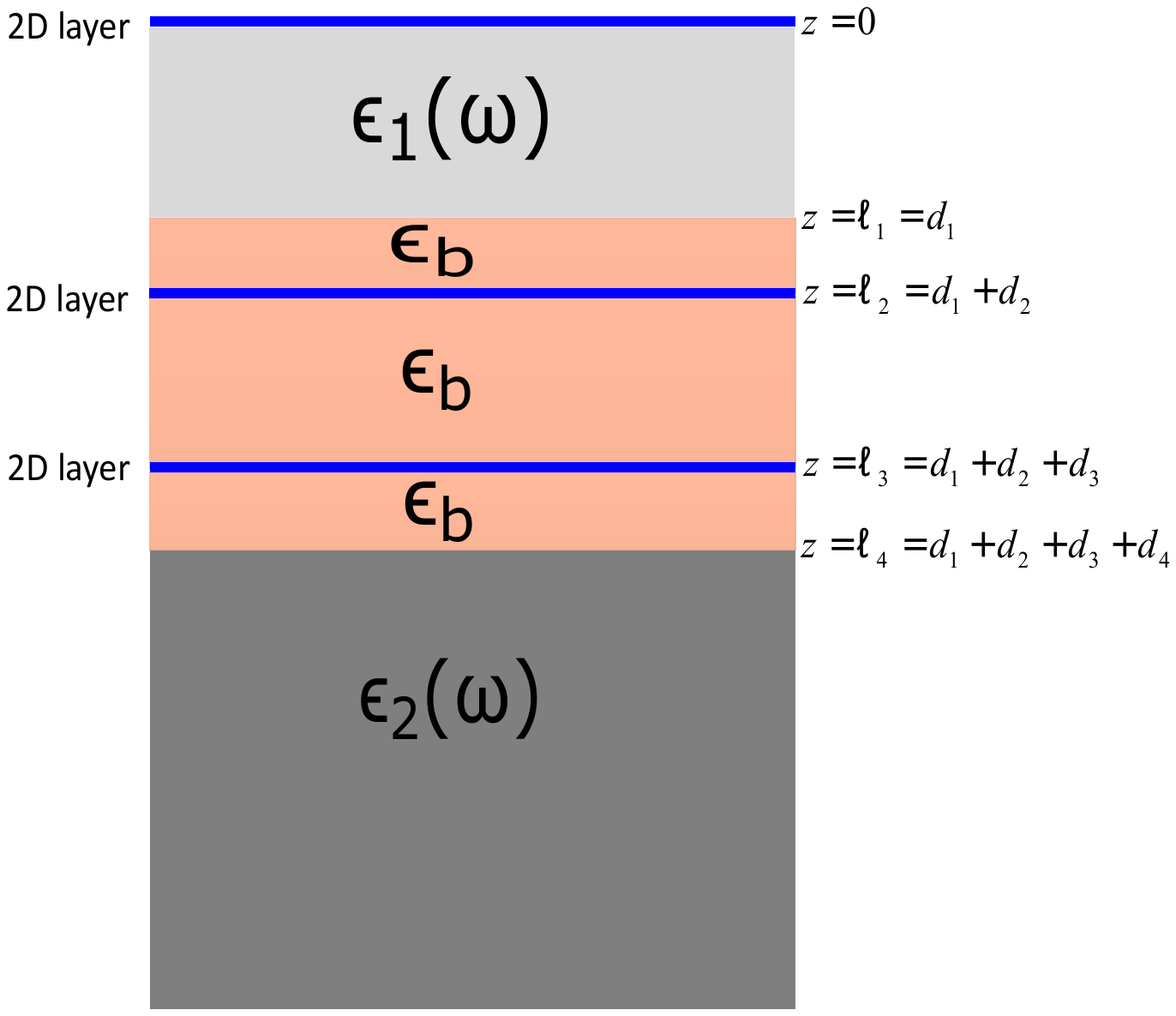}
\caption{(Color online)Schematic illustration of a pair of 2d layer at $z=\ell_2$ and at $z=\ell_3$ embedded in the background dielectric material of dielectric constant $\epsilon_b$ and encapsulated by conducting substrate of dielectric function $\epsilon_1(\omega)$ and $\epsilon_2(\omega)$. A protective coat of 2d layer lies on top of this heterostructure at $z =0$.} 
\label{FIG:1}
\end{figure}

\begin{figure}[t]
\centering
\includegraphics[width=7.4in,height=5.1in,keepaspectratio]{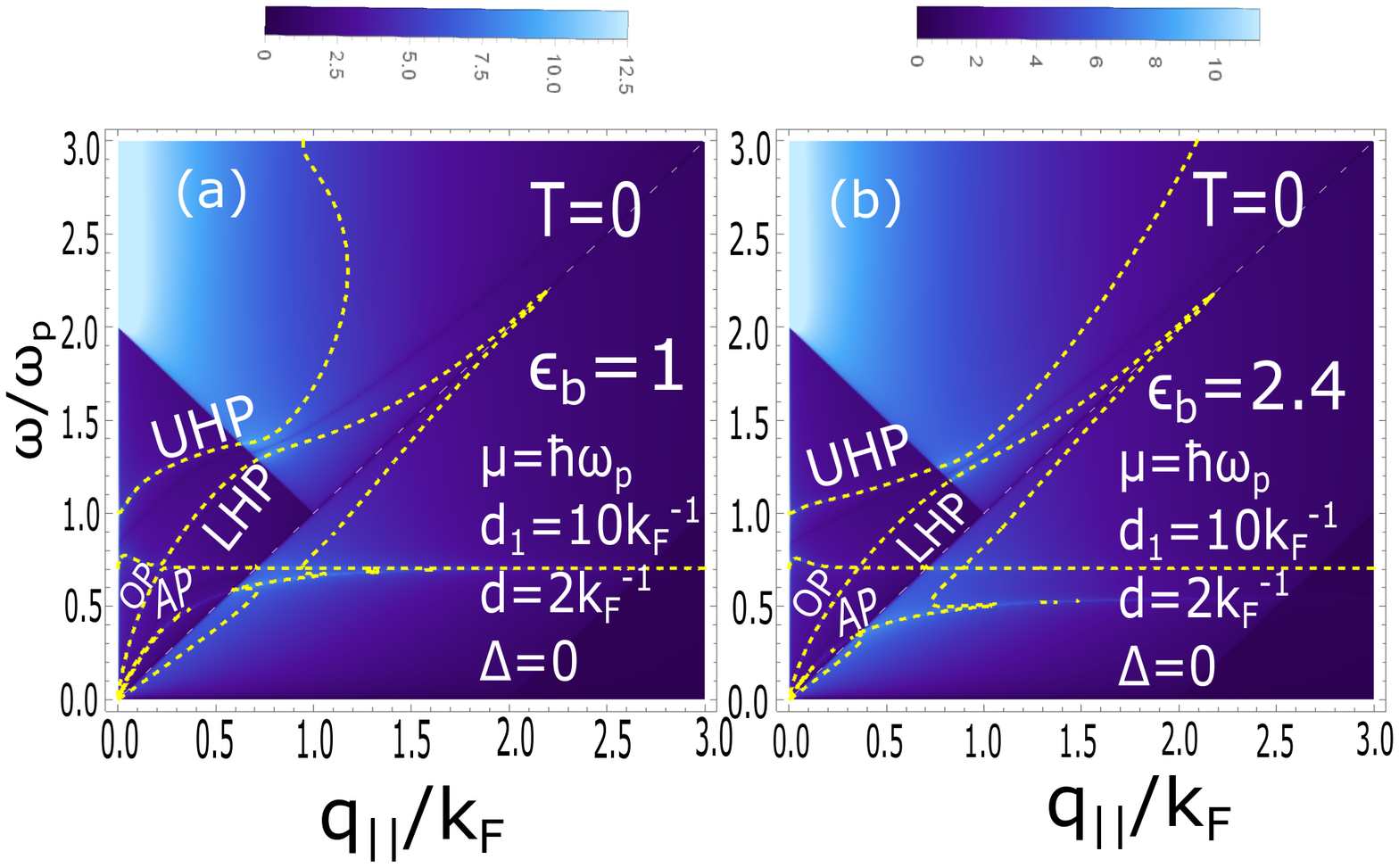}
\caption{(Color online)  Panel(a) and (b) present density plots at zero temperature of the plasmon dispersion relation for 
a double layer {\em gapless}, doped graphene structure inserted in (a) vacuum with  background dielectric constant $\epsilon_b=1$ 
and (b) dielectric with background  dielectric constant $\epsilon_b=2.4$ and sandwiched between two conducting plasmas, one 
with finite thickness and another which is semi-infinite. The dashed lines in the dark regions show the 
plasma resonance for chosen values of the chemical potential $\mu$, dielectric thickness $d_1$ and inter-layer 
separation $d$.   Also, $k_F$ is the Fermi wave vector for graphene. An acoustic plasmon (AP), optical plasmon (OP)
and two high-frequency modes labeled as a lower hybrid plasmon (LHP) and upper hybrid plasmon (UHP)
also appear in the excitation spectrum. }
\label{FIG:2}
\end{figure}

\begin{figure}[t]
\centering
\includegraphics[width=8.6in,height=6.1in,keepaspectratio]{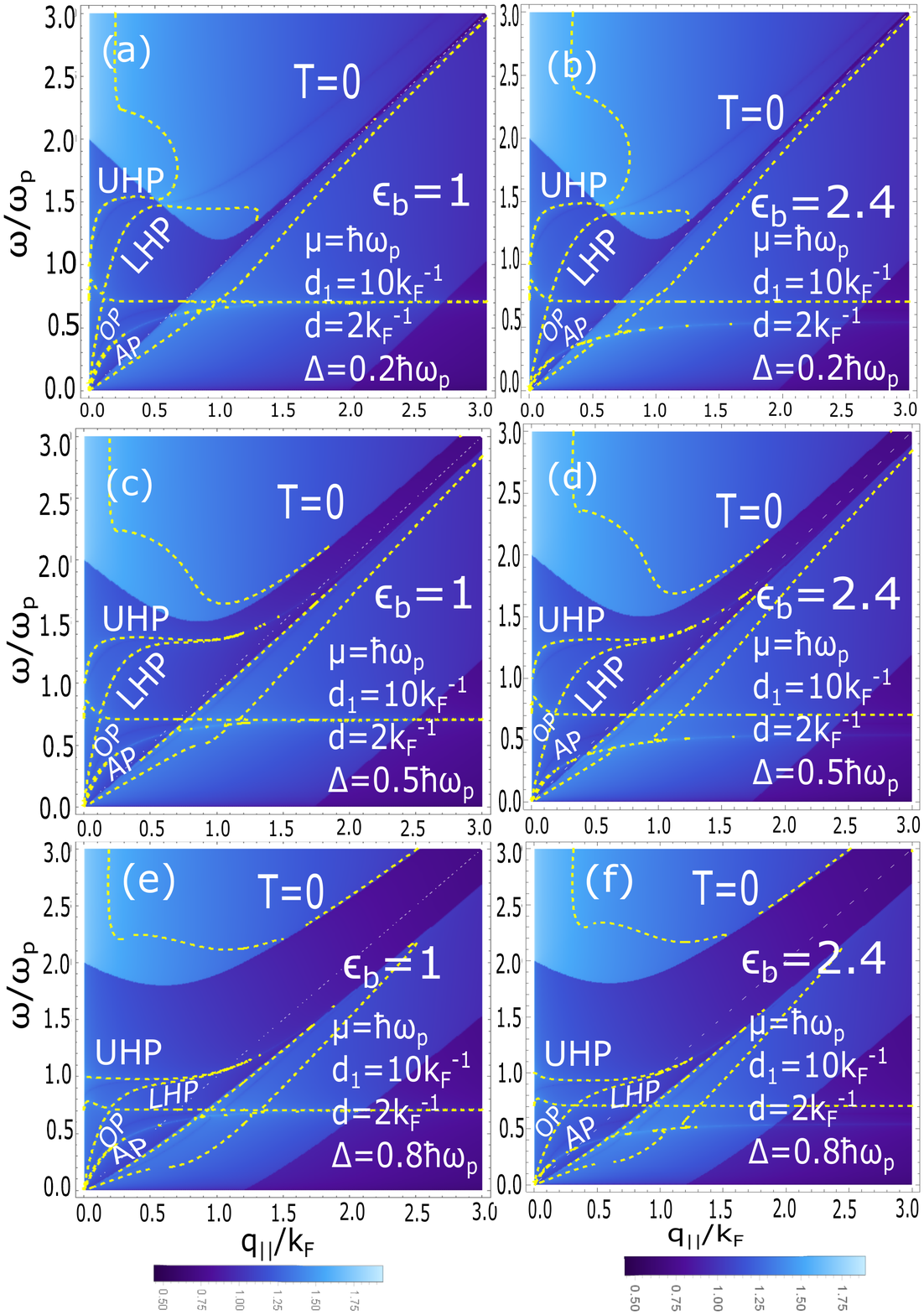}
\caption{(Color online) The same as in Fig. \ref{FIG:2}, except that the half-band energy gap is {\em finite}.
In (a) and (b), $\Delta$=$0.2\hbar \omega_p$, for (c) and (d)$\Delta$=$0.5\hbar \omega_p$ and for (e) and (f) $\Delta$=$0.8 \hbar \omega_p$. 
The 2D layers lie in vacuum for the  left panels and in background dielectric constant $\epsilon_b=2.4$ for right panels. }
\label{FIG:3}
\end{figure}

\begin{figure}[t]
\centering
\includegraphics[width=8.6in,height=6.1in,keepaspectratio]{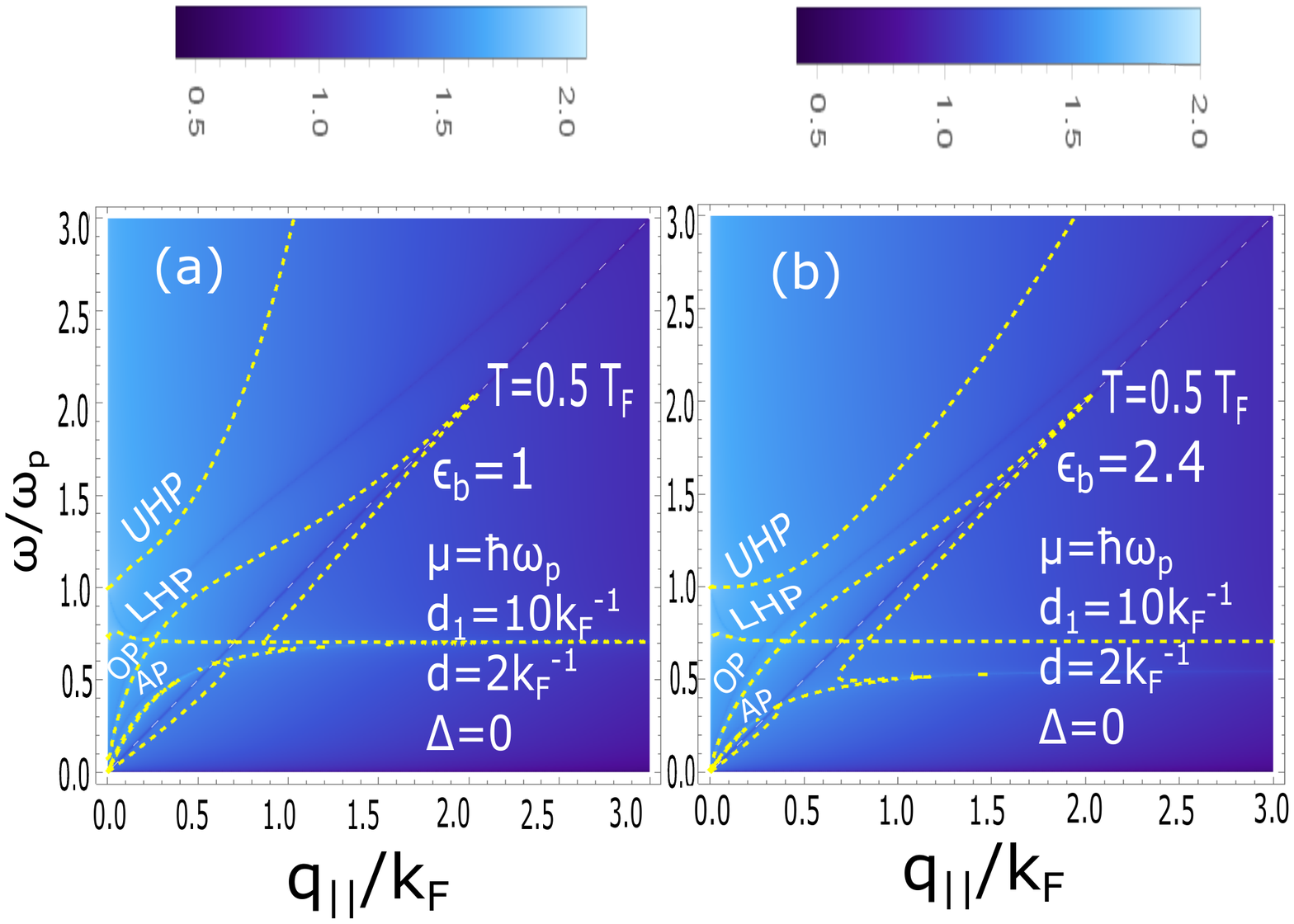}
\caption{(Color online) The same as in Fig.\ \ref{FIG:2}. except that  $T=0.5T_F$ for a pair of graphene sheets lying in vacuum(left panel) 
and in background dielectric constant $\epsilon_b=2.4$(right panel). }
\label{FIG:4}
\end{figure}

We have presented a formalism for calculating the surface response function  for a layered 
2D structure which is sandwiched by conducting materials. The 2D layers may be separated by
dielectric materials with arbitrary thickness.  Our formulation uses a transfer matrix method for
explicitly calculating the electrostatic potentials and electric fields along with linear response theory for the induced
charge density on the 2D layers.  We could include the effect on the surface response function 
due to a 2D layer on the surface of the hybrid structure as illustrated in Fig.~\ref{FIG:1}. 
However, the presence of this 2D layer was neglected in our calculations.  For comparison, we have 
presented numerical results when the encapsulated 2D layers are not embedded in a dielectric with that
when these layers have a dielectric material on either side. Our results are shown in Figs.~\ref{FIG:2} and 
\ref{FIG:3} for gapless and gapped graphene, respectively. The effect due to temperature on these self-sustaining
modes  appears in our presented plots in Fig.~\ref{FIG:4}.

 \medskip
\par

 These plasmons may play an important role in fundamental studies involving
  strong  light-matter interactions on the nanoscale. Furthermore, the existence of
a branch with   acoustic   dispersion could offer   many-fold novel possibilities
for the development of devices for detector, sensor and communication applications
in the technologically important THz range, such as nanoscale waveguides or modulators.

\medskip
\par

We have demonstrated that we may tune the plasma frequency of a
double layer graphene heterostructure by adjusting the doping
concentrations. Our model calculations show that these devices
have potential  for high-frequency operation and large-scale
integration.  We only consider high doping concentrations so that
the Fermi level is far away from the Dirac {\bf K} point.  Otherwise,
localization effects on the charge carriers cannot be neglected.
As a matter of fact, the broken electron-hole symmetry may be  attributed to the  mutual
 polarization of the closely spaced interacting layers and impurity scattering.
The encapsulated double-layer graphene plasmons we are predicting possess
 strong field confinement and very low damping.  This enables  a new class of devices
 for tunable subwavelength heterostructures,   strong light-matter interactions and
  nano-optoelectronic switches. Although all of these   prospects require low plasmon
  damping, our model calculations show that this may be achieved  if the Fermi level
  is not too low  so that  impurity scattering may be neglected.

\appendix

\medskip
\par

\newpage

\end{document}